\documentclass[12pt]{amsart}
\usepackage{amsmath,amssymb,cite,enumerate,graphicx,hyperref}

\hypersetup{pdftex,colorlinks=true,allcolors=blue}
\usepackage[all]{hypcap}

\usepackage[foot]{amsaddr}
\usepackage[sc]{mathpazo}

\newcommand\ack{\subsection*{Acknowledgment}}

\DeclareMathAlphabet\mathsfbi{T1}{phv}{b}{it}

\numberwithin{equation}{section}

\newcommand\BV{\boldsymbol} 
\newcommand\BM{\mathsfbi} 

\newcommand\dif{\mathrm{d}}
\newcommand\deriv[2]{\frac{\dif #1}{\dif #2}}
\newcommand\parderiv[2]{\frac{\partial #1}{\partial #2}}
\newcommand\coll{\mathcal C}

\newcommand\EE{\mathbb E}
\newcommand\trace{\mathrm{tr}}

\newcommand\eqdef{\stackrel{\text{def}}=}

\newcommand\Pran{\mbox{\textit{Pr}}}
\newcommand\Schm{\mbox{\textit{Sc}}}
\newcommand\im{\nu}

\newcommand\cL{\mathcal L}
\newcommand\cN{\mathcal N}
\newcommand\cW{\mathcal W}
\newcommand\cE{\mathcal E}

\newcommand\myatop[2]{\genfrac{}{}{0pt}{}{#1}{#2}}

\begin{document}

\author[Rafail V. Abramov]{Rafail V. Abramov}

\address{Department of Mathematics, Statistics and Computer Science,
University of Illinois at Chicago, 851 S. Morgan st., Chicago, IL 60607}

\email{abramov@uic.edu}

\title{A mass diffusion effect in gas dynamics equations}

\begin{abstract}
In the current work we propose a theory for an additional mass
diffusion effect in the conventional gas dynamics equations. We find
that this effect appears as a homogenization time limit correction,
when the deterministic interaction process of the real gas molecules
is replaced with a simplified random interaction process for
consistency with the Boltzmann equation. For the simplified random
interaction processes represented by either a hard sphere random
scattering model, or by a model which employs the Lennard-Jones
potential for random molecular deflections, we compute the estimates
of the corrective diffusion coefficient in the Euler, Navier-Stokes
and Grad equations for some monatomic and polyatomic gases.
\end{abstract}



\maketitle

\section{Introduction}

Recently, there has been a profound interest in the extended fluid
dynamics \cite{Brenner,Brenner2,Brenner3,Durst,Durst2} and the
underlying Boltzmann equation \cite{DadRee,Kli,Sher}, which included
an additional spatial diffusion term. A concept of the ``volume
velocity'' was introduced in \cite{Brenner,Brenner2,Brenner3}, which
differed from the conventional mass velocity by a small flux term
derived from Fick's law. Similar additional terms to model the
self-diffusion of mass were introduced in \cite{Durst,Durst2}. In
\cite{DadRee,Kli,Sher}, an \textit{ad hoc} diffusion term was
introduced directly into the Boltzmann equation. Also,
in~\cite{GreRee}, it was demonstrated that an additional mass
diffusion term in the Navier-Stokes equations provided a more accurate
model for a supersonic shock wave.

In our recent work~\cite{Abr13}, we found that a simplified model of
the gas molecule interactions based on the appropriately chosen random
jump process is more consistent with the Boltzmann
equation~\cite{Cer2,CerIllPul} than the deterministic multimolecular
dynamics of a real gas based on a repelling potential (such as the
Lennard-Jones potential~\cite{Len}). We then speculated that a
difference between the two models could manifest in the form of a
corrective mass diffusion term with an empirically determined mass
diffusion coefficient $D_\alpha$. Also, in~\cite{Abr13} we tested the
proposed mass diffusion model in the Couette microflow setting with
the appropriately scaled near-wall viscosity~\cite{Abr15} and fully
recovered the basic moments of the benchmark Direct Simulation Monte
Carlo (DSMC) \cite{Bird,ScaRooWhiDarRee} computation in the same
setting, including the Knudsen layers, for the diffusive Navier-Stokes
and regularized Grad equations. In the subsequent work~\cite{AbrOtt}
we tested the same diffusive model of~\cite{Abr13} with the same
viscosity scaling~\cite{Abr15} in the Poiseuille microflow setting,
and recovered the basic moments of the corresponding benchmark DSMC
computation with high accuracy (about 1\% relative error in the
velocity, mass flow, temperature and pressure).

In the current work, we propose a more coherent theory for the mass
diffusion effect in the gas dynamics equations, based on a diffusion
correction of the simplified random molecule interaction process to
match the dynamics of a realistic gas in the homogenization time
limit~\cite{KelMel,Kif,PavStu,Van}. For the corresponding simplified
interaction model of the gas molecules based on randomly colliding
hard spheres~\cite{ChaCow,Dym,Hey} and random deflections via the
Lennard-Jones potential~\cite{Len,HirBirSpo,HirBirSpo2,Winn}, we
subsequently compute the estimates of the corrective mass diffusion
coefficient $D_\alpha$ in the Euler, Navier-Stokes~\cite{Bat,Gols} and
Grad~\cite{Gra,Mal,StruTor} equations for some monatomic and
polyatomic gases.

The manuscript is organized as follows. In Section~\ref{sec:av_hom} we
review the multiscale time approach~\cite{KelMel,Kif,PavStu,Van} for a
general multiparticle interaction system~\cite{Abr13}, while also
including an external acceleration (such as, for example, the
planetary gravity) which may optionally affect the gas molecules. In
Section~\ref{sec:diff_correction} we introduce a multiparticle system
with a simplified random molecular interaction process with an
additional diffusion correction to ensure consistency with the
realistic gas dynamics in the homogenization time limit. In
Section~\ref{sec:boltzmann} we review the random jump molecular
interaction operator from~\cite{Abr13} and the derivation of the
corresponding diffusive Boltzmann equation, which now includes the
optional external acceleration. In Sections~\ref{sec:Euler},
\ref{sec:Navier-Stokes} and~\ref{sec:Grad} we review, respectively,
the diffusive Euler, Navier-Stokes and Grad equations, and compute the
estimates of the corrective mass diffusion coefficients for neon,
argon, krypton, xenon, nitrogen, oxygen, carbon dioxide and
methane. We discuss the results of the work in
Section~\ref{sec:summary}.

\section{Averaging and homogenization for a fast interaction dynamics}
\label{sec:av_hom}

Here we review the homogenization formalism for a multiparticle gas
system we introduced in~\cite{Abr13}, and equip it with an optional
external acceleration term (which could be important in practical
applications). The relevant theory can be found
in~\cite{KelMel,Kif,PavStu,Van}.

Consider a general system of $K$ particles in an $N$-dimensional
space, where the $i$-th particle has the coordinate $\BV x_i$ and
velocity $\BV v_i$. For the whole system of particles, we concatenate
their respective coordinate and velocity vectors into
\[
\BV X=(\BV x_1,\ldots,\BV x_K),\qquad\BV V=(\BV v_1,\ldots,\BV v_K).
\]
For realistic gases, the first 3 components of each $\BV x_i$ or $\BV
v_i$ refer to the translational movement of molecules, while the
remaining $(N-3)$ components refer to rotational degrees of freedom,
and thus are naturally periodic. In what is presented below, we assume
that the translational coordinates are also periodic (that is, the
3-dimensional physical space is folded into a torus), to keep all $K$
molecules in a finite volume and avoid boundary effects at the same
time.

We start with a general $(\BV X(t),\BV V(t))$-process,
given by
\begin{equation}
\label{eq:dyn_sys}
\deriv{\BV X}t=\BV V,\qquad\dif\BV V=\BV A\dif t+\dif\BV\cN(t,\BV
X,\BV V).
\end{equation}
Above,  $\BV A$ is an external
acceleration of the form
\begin{equation}
\BV A=(\BV a_1,\ldots\BV a_K),
\end{equation}
where $\BV a_i$ is an individual external acceleration which acts on
the $i$-th molecule. We assume that only the first 3 components of any
$\BV a_i$ are nonzero (that is, the external acceleration may only act
along the translational directions). Also, $\BV\cN$ is a general,
possibly random, stationary interaction process between the
molecules. We assume that $\BV\cN$ preserves the momentum $\BV m$ and
energy $\cE$ of the form
\begin{equation}
\label{eq:m_E}
\BV m=\sum_{k=1}^K\BV v_k,\qquad\cE=\frac 12\sum_{k=1}^K\|\BV v_k\|^2
=\frac 12\|\BV V\|^2,
\end{equation}
where $\|\cdot\|$ is the usual Euclidean norm. Note that for realistic
gas interaction processes the energy does not have a purely kinetic
form above, as it normally also includes the interaction potential
(such as the Lennard-Jones potential~\cite{Len}). We will, however,
assume that the gas is dilute, that is, the effective range of the
potential is negligible in comparison with the average distance
between the molecules, so that most of the time the energy indeed
consists solely of its kinetic part in~\eqref{eq:m_E}.

The corresponding Kolmogorov equation for the process
in~\eqref{eq:dyn_sys} is, subsequently, given by
\begin{equation}
\label{eq:kolmogorov}
\parderiv Ft+\nabla_{\BV V}\cdot(\BV AF)=\cL^* F,
\end{equation}
where $\cL$ is the infinitesimal generator of~\eqref{eq:dyn_sys} with
the external acceleration $\BV A$ set to zero. The general form of
$\cL$ is given by
\begin{equation}
\cL=\BV V\cdot\nabla_{\BV X} + \{\text{velocity interaction}\},
\end{equation}
where the second term in the right-hand side is given through the
properties of $\BV\cN$ (and thus is undefined, for now). We will
assume that any solution $(\BV X(t),\BV V(t))$ of~\eqref{eq:dyn_sys}
with $\BV A$ set to zero is ergodic and strongly mixing on the
corresponding union of $\BV X$-torus and $\BV V$-sphere. Thus, both
$\cL$ and its adjoint $\cL^*$ possess a one-dimensional null space of
constant functions, such that, in the absence of the external
acceleration $\BV A$, any starting state of $F$
in~\eqref{eq:kolmogorov} is forced by $\cL^*$ to relax, with time,
towards the uniform invariant state on the $\BV X$-torus and $\BV
V$-sphere.

\subsection{Averaging for the fast interaction dynamics}

Let us rescale the dynamical system in~\eqref{eq:dyn_sys} as follows:
\begin{equation}
\label{eq:dyn_sys_rescaled}
\deriv{\BV X}t=\BV V,\qquad\deriv{\BV Y}t=\frac 1\varepsilon\BV V,
\qquad\dif\BV V=\BV A\dif t+\dif\BV\cN(t/\varepsilon,\BV Y,\BV V),
\end{equation}
where $0<\varepsilon\ll 1$ is a small constant scaling parameter.
Here we assume that $\BV\cN$ is the ``fast'' process (as compared to
the external acceleration $\BV A$), and we also introduce the ``fast''
coordinate process $\BV Y$, which formally replaces $\BV X$ (as the
latter becomes the ``slow'' coordinate process). This rescaling is
similar to what was done, for example, in Section 11.7.5
of~\cite{PavStu}. Observe that the rescaled system
in~\eqref{eq:dyn_sys_rescaled} is ``backward-compatible'' with the
original one in~\eqref{eq:dyn_sys} if one sets $\varepsilon=1$. The
corresponding rescaled Kolmogorov equation is then given by
\begin{equation}
\label{eq:kolmogorov_rescaled}
\parderiv Ft+\BV V\cdot\nabla_{\BV X}F+\nabla_{\BV V}\cdot(\BV AF) =
\frac 1\varepsilon\cL^*F,
\end{equation}
where $\cL^*$ now depends on $\BV Y$, rather than on $\BV X$. We now
expand $F$ in powers of $\varepsilon$ as
\begin{equation}
\label{eq:F_expansion}
F=F_0+\varepsilon F_1+\varepsilon^2 F_2+\ldots,
\end{equation}
where each $F_i$ does not depend on $\varepsilon$. This yields, for
the successive powers of $\varepsilon$,
\begin{equation}
\cL^*F_0=0,\qquad\parderiv{F_0}t+\BV V\cdot\nabla_{\BV
  X}F_0+\nabla_{\BV V}\cdot(\BV A F_0)=\cL^*F_1.
\end{equation}
From the first identity above we find that $F_0$ is an invariant
measure for the process $(\BV Y,\BV V)$ in the absence of the external
acceleration $\BV A$. In particular, let
\begin{equation}
\bar F=\int F\,\dif\BV Y\dif\BV V,
\end{equation}
then for any suitable $\psi$ we have
\begin{equation}
\int\psi F_0\,\dif\BV Y\dif\BV V=\bar F_0\int\psi\,\dif\im,
\end{equation}
where $\im$ is the uniform invariant probability measure on the $\BV
Y$-torus and on the constant momentum/energy sphere of $\BV V$. Then,
integrating the second identity above over $\dif\BV Y\dif\BV V$, we
find
\begin{equation}
\label{eq:kolmogorov_averaged}
\parderiv{\bar F_0}t+\nabla_{\BV X}\cdot(\bar F_0\BV U_0)=0,
\end{equation}
where $\BV U_0$ is the average velocity of the fast system:
\begin{equation}
\BV U_0=\int\BV V\,\dif\im.
\end{equation}
Due to the fact that all particles are statistically identical when
$\BV V$ is uniformly distributed on the constant momentum/energy
sphere, $\BV U_0$ consists of $K$ identical vectors $\BV u_0$:
\begin{equation}
\BV U_0=(\BV u_0,\ldots,\BV u_0),
\end{equation}
and can be set to zero via an appropriate Galilean shift of the
coordinate system in the $N$-dimensional space where the
multimolecular system in~\eqref{eq:dyn_sys} evolves.

\subsection{Homogenization for the fast interaction dynamics}

We are now going to set the average velocity $\BV u_0$ of the fast
system above to zero via an appropriate Galilean shift, which sets
$\partial\bar F_0/\partial t=0$ at the $O(1)$ time scale, according
to~\eqref{eq:kolmogorov_averaged}. This, however, simply means that
the time derivative of $F$ manifests on the longer time scale
$O(\varepsilon^{-1})$~\cite{PavStu,Van}, so we rescale
\begin{equation}
\label{eq:F_rescale}
\parderiv Ft\to\varepsilon\parderiv Ft
\end{equation}
above in~\eqref{eq:kolmogorov_rescaled}. In this situation, the
expansion in~\eqref{eq:F_expansion} yields
\begin{subequations}
\begin{equation}
\cL^*F_0=0,\qquad\cL^*F_1=\BV V\cdot\nabla_{\BV X}F_0+\nabla_{\BV V}
\cdot(\BV AF_0),
\end{equation}
\begin{equation}
\parderiv{F_0}t+\BV V\cdot\nabla_{\BV X}F_1+\nabla_{\BV V}\cdot(\BV A
F_1)=\cL^*F_2.
\end{equation}
\end{subequations}
Apparently, $F_0$ has the same meaning as above in the averaging case,
and integrating the third equation in $\dif\BV Y\dif\BV V$ yields
\begin{equation}
\parderiv{\bar F_0}t+\nabla_{\BV X}\cdot\left(\int \BV VF_1\,\dif\BV
Y\dif\BV V\right)=0.
\end{equation}
To compute the integral above, let us define the process~$\Phi(t,\BV
Y,\BV V)$ as
\begin{equation}
\Phi(t,\BV Y,\BV V)=\int_0^t\EE\BV V(s)\dif s.
\end{equation}
where the notation $\EE\BV V(t)$ is a shorthand for
\begin{equation}
\EE\BV V(t)\eqdef\EE[\BV V(t)|\BV Y(0)=\BV Y,\BV V(0)=\BV V],
\end{equation}
where the velocity process $\BV V(t)$ is generated
by~\eqref{eq:dyn_sys} with $\BV A$ set to zero. Above, the expectation
$\EE$ is included in case $\BV\cN$ happens to be a random
process. Now, by the definition of the generator~$\cL$,
\begin{equation}
\cL\Phi(t,\BV Y,\BV V)=\lim_{h\to 0} \frac{ \EE \Phi(t,\BV Y(h),\BV
  V(h))-\Phi(t,\BV Y,\BV V)}h=\EE\BV V(t)-\BV V.
\end{equation}
For more details on the above identity, see, for example, Lemma 3.2.2
in~\cite{App}. Now, let us consider the integral of the form
\begin{multline}
\int\BV V F_1\,\dif\BV Y\dif\BV V-\int\EE\BV V(t) F_1\,\dif\BV Y\dif
\BV V=-\int\cL\Phi(t,\BV Y,\BV V) F_1\,\dif\BV Y\dif\BV V=\\=-\int
\Phi(t,\BV Y,\BV V)\cL^* F_1\,\dif\BV Y\dif\BV V=-\int_0^t\dif s
\int\EE\BV V(s) \Big(\BV V\cdot\nabla_{\BV X}F_0+\nabla_{\BV V}\cdot
(\BV AF_0)\Big) \,\dif\BV Y\dif\BV V.
\end{multline}
Observe that the first term in the first line is what we need to
compute, while the last identity depends entirely on the properties of
the leading-order dynamics. Thus, in order to express the former in
terms of the latter, we need to eliminate the second term in the first
line, where $F_1$ depends on $\BV Y$ and $\BV V$ (but not on $\BV
V(t)$), and we are also at freedom to choose the time parameter $t$.

First, let us note that the integration over $\dif\BV Y\dif\BV V$
above occurs over the $\BV Y$-torus and $\BV V$-sphere (with the
latter now centered at zero momentum due to the choice of the
coordinates). Thus, we can write
\begin{equation}
\int\EE\BV V(t)F_1\,\dif\BV Y\dif\BV V\sim\int\EE\BV V(t)F_1 \,
\dif\im,
\end{equation}
where $\im$ is, as before, the uniform probability measure of the fast
interaction dynamics, and the constant proportionality coefficient
involves the surface areas of the $\BV Y$-torus and $\BV V$-sphere.
Next, the strong mixing assumption for~\eqref{eq:dyn_sys} (and,
equivalently, \eqref{eq:dyn_sys_rescaled}) yields
\begin{equation}
\lim_{t\to\infty}\int\EE\BV V(t) F_1\,\dif\im=\BV 0,
\end{equation}
and, therefore,
\begin{equation}
\int\BV V F_1\,\dif\BV Y\dif\BV V=-\int_0^\infty\dif s\int\EE\BV V(s)
\Big(\BV V\cdot\nabla_{\BV X} F_0+\nabla_{\BV V}\cdot(\BV AF_0)\Big)
\,\dif\BV Y\dif\BV V.
\end{equation}
The expression above is still practically intractable, a thus a
further simplification is needed. As in~\cite{Abr13}, here we are
going to assume that the statistics of the velocity process exhibit
properties similar to Gaussian random processes. Namely, we are going
to assume that the following analog of the regression
theorem~\cite{Ris} holds for a suitable $\psi(\BV Y,\BV V)$:
\begin{equation}
\int\EE\BV V(s)\psi(\BV Y,\BV V) \,\dif\BV Y\dif\BV V=\BM C(s)\BM
C^{-1}(0)\int\BV V\psi(\BV Y,\BV V) \,\dif\BV Y\dif\BV V.
\end{equation}
Above, $\BM C(t)$ is the velocity time autocorrelation function
\begin{equation}
\BM C(t)=\int\EE\BV V(t)\BV V^T\dif\im.
\end{equation}
This assumption yields
\begin{equation}
\int\BV V F_1\,\dif\BV Y\dif\BV V=-\left(\int_0^\infty\BM C(s)\dif
s\right)\BM C^{-1}(0)\int\BV V\Big(\BV V\cdot\nabla_{\BV X}
F_0+\nabla_{\BV V}\cdot(\BV AF_0)\Big) \,\dif\BV Y\dif\BV V.
\end{equation}
Now observe that
\begin{subequations}
\begin{equation}
\int\BV V\Big(\BV V\cdot\nabla_{\BV X} F_0\Big)\,\dif\BV Y\dif\BV V=
\nabla_{\BV X}\cdot\int\BV V\BV V^TF_0\,\dif\BV Y\dif\BV V=\nabla_{\BV
  X}\cdot\left(\bar F_0\int\BV V\BV V^T\dif\im\right)=\nabla_{\BV
  X}\cdot\left(\bar F_0\BM C(0)\right),
\end{equation}
\begin{equation}
\int\BV V\Big(\nabla_{\BV V}\cdot(\BV AF_0)\Big) \,\dif\BV Y\dif\BV
V=-\int\BM I\BV AF_0\,\dif\BV Y\dif\BV V=-\bar F_0\int\BV
A\,\dif\im=-\bar F_0\BV{\bar A}_0,
\end{equation}
\end{subequations}
where $\BV{\bar A}_0$ is the $\im$-average of $\BV A$. Thus, we
further obtain
\begin{equation}
\int\BV V F_1\,\dif\BV Y\dif\BV V=-\left(\int_0^\infty\BM C(s)\dif
s\right)\BM C^{-1}(0)\Big(\nabla_{\BV X}\cdot\left(\bar F_0\BM
C(0)\right)-\bar F_0\BV{\bar A}_0\Big)
\end{equation}
Even in this simplified form, the velocity autocorrelation function
$\BM C(t)$ and its inverse are $KN\times KN$ matrices, and thus are
still computationally intractable. Thus, as in~\cite{Abr13}, we make
yet another simplifying assumption:
\begin{equation}
\BM C(0)=\theta_0\BM I,\qquad\left(\int_0^\infty\BM C(s)\,\dif
s\right)\BM C^{-1}(0)=\tau_0\BM I.
\end{equation}
This can be justified by recalling that all particles in the fast
interaction process are statistically identical, and that their number
is large enough to make them statistically decorrelated (even though,
strictly speaking, they are not independent since they reside on a
common constant energy sphere). The quantities $\theta_0$
(temperature) and $\tau_0$ (decorrelation time) are thus given by
\begin{equation}
\theta_0=\frac 1{KN}\int \|\BV V\|^2\dif\im,\qquad\tau_0=\frac
1{KN\theta_0}\int_0^\infty \dif s\int\EE\BV V(s)\cdot\BV V(0)\dif\im.
\end{equation}
The latter simplification finally yields a tractable homogenization
time limit
\begin{equation}
\label{eq:kolmogorov_homogenization}
\parderiv{\bar F_0}t=\nabla_{\BV X}\cdot\left(\tau_0\left(\nabla_{\BV
  X}\left(\theta_0\bar F_0\right)-\BV{\bar A}_0\bar F_0\right)\right).
\end{equation}

\section{A diffusion correction for simplified interaction processes}
\label{sec:diff_correction}

Let us assume that the interaction process $\BV\cN$
in~\eqref{eq:dyn_sys} was that of a realistic gas, and that it can be
approximated by a different, simplified interaction process
$\BV\cN_s$, with the corresponding simplified molecular dynamics
equations accordingly given by
\begin{equation}
\label{eq:dyn_sys_s}
\deriv{\BV X}t=\BV V,\qquad\dif\BV V=\BV A\dif t+\dif\BV\cN_s(t,\BV
X,\BV V).
\end{equation}
We will assume that the simplified interaction process $\BV\cN_s$ also
preserves the momentum and energy in~\eqref{eq:m_E} and has the same
invariant uniform probability measure $\im$ on the sphere of constant
momentum and energy, so that the momentum and energy (and, therefore,
the fast temperature $\theta_0$) of $\BV\cN_s$ match those of
$\BV\cN$. As before, we choose a suitable coordinate system via an
appropriate Galilean shift so that the momenta of both~$\BV\cN$ and
$\BV\cN_s$ equal zero.

Observe that while the system in~\eqref{eq:dyn_sys_s} is a valid
approximation of~\eqref{eq:dyn_sys} in the averaging time limit (that
is, on the $O(1)$ time scale), their corresponding homogenization time
limits differ due to the difference in the velocity
autocorrelations. Here we propose a more accurate modification of the
simplified interaction system in~\eqref{eq:dyn_sys_s}, which is
consistent with~\eqref{eq:dyn_sys} in the homogenization time limit.

Let us first look at the following two rescaled systems:
\begin{subequations}
\begin{equation}
\label{eq:dyn_sys_slow}
\deriv{\BV Y}t=\BV V,\qquad\dif\BV V=\varepsilon\BV A\dif t+\dif
\BV\cN(t,\BV Y,\BV V),
\end{equation}
\begin{equation}
\label{eq:dyn_sys_s_slow}
\deriv{\BV Y}t=\BV V,\qquad\dif\BV V=\varepsilon\BV A\dif t+\dif
\BV\cN_s(t,\BV Y,\BV V).
\end{equation}
\end{subequations}
Observe that these two processes are the analogs
of~\eqref{eq:dyn_sys_rescaled}, except that the time scale of~$\BV\cN$
and $\BV\cN_s$ was not accelerated, but rather the one of the external
acceleration $\BV A$ was slowed down instead. Let $\tau_\varepsilon$
refer to the decorrelation time of~\eqref{eq:dyn_sys_slow}, and let
$\tau_{s,\varepsilon}$ be the same for~\eqref{eq:dyn_sys_s_slow}. The
temperature $\theta_\varepsilon$ and the averaged acceleration
$\BV{\bar A}_\varepsilon$ will refer to the simplified system
in~\eqref{eq:dyn_sys_s_slow}. Clearly, with $\varepsilon$ set to zero
we arrive to the limiting quantities $\theta_0$ and $\tau_0$ of the
previous section for both realistic and simplified systems; on the
other hand, setting $\varepsilon=1$ yields the corresponding
temperature and decorrelation times of~\eqref{eq:dyn_sys}
and~\eqref{eq:dyn_sys_s} (and in this case we drop the subscript, that
is, $\theta_1=\theta$, $\BV{\bar A}_1=\BV{\bar A}$, $\tau_1=\tau$, and
$\tau_{s,1}=\tau_s$).

Now, consider the following It\^o stochastic differential equation,
which is also based on the simplified fast interaction operator
$\BV\cN_s$:
\begin{subequations}
\label{eq:dyn_sys_approx}
\begin{equation}
\dif\BV X=\big(\BV V+\theta\nabla_{\BV X}\left(\tau-\tau_s\right)+
(\tau-\tau_s)\BV{\bar A}\big)\dif t+\sqrt{2\theta(\tau-\tau_s)}
\dif\BV\cW(t),
\end{equation}
\begin{equation}
\dif\BV V=\BV A\dif t+\dif\BV\cN_s(t,\BV X,\BV V),
\end{equation}
\end{subequations}
where $\BV\cW(t)$ is a $KN$-dimensional Wiener process, while
$\theta$, $\BV{\bar A}$, $\tau$ and $\tau_s$ are those
of~\eqref{eq:dyn_sys_slow} and~\eqref{eq:dyn_sys_s_slow}, with
$\varepsilon=1$ as described above. We are now going to show that,
when rescaled in the same manner as above
in~\eqref{eq:dyn_sys_rescaled}, the system
in~\eqref{eq:dyn_sys_approx} has the same homogenization limit as the
one in the realistic system~\eqref{eq:dyn_sys}, despite the fact that
the interactions of particles are governed by the simplified
process~$\BV\cN_s$, rather than by the realistic
process~$\BV\cN$. Indeed, with the same rescaling as
in~\eqref{eq:dyn_sys_rescaled}, we have
\begin{subequations}
\label{eq:dyn_sys_approx_rescaled}
\begin{equation}
\dif\BV X=\left(\BV V+\varepsilon\theta_\varepsilon\nabla_{\BV X}
\left(\tau_\varepsilon-\tau_{s,\varepsilon}\right)+\varepsilon
(\tau_\varepsilon-\tau_{s,\varepsilon})\BV{\bar A}_\varepsilon\right)
\dif t+\sqrt{2\varepsilon\theta_\varepsilon(\tau_\varepsilon
  -\tau_{s,\varepsilon})}\dif\BV\cW(t),
\end{equation}
\begin{equation}
\dif\BV Y=\left(\frac 1\varepsilon\BV V+\theta_\varepsilon\nabla_{\BV
  Y}\left(\tau_\varepsilon-\tau_{s,\varepsilon}\right)+
(\tau_\varepsilon-\tau_{s,\varepsilon})\BV{\bar A}_\varepsilon\right)
\dif t+\sqrt{2\theta_\varepsilon(\tau_\varepsilon
  -\tau_{s,\varepsilon}) }\dif\BV\cW(t),
\end{equation}
\begin{equation}
\dif\BV V=\BV A\dif t+\dif\BV\cN_s(t/\varepsilon,\BV Y,\BV V).
\end{equation}
\end{subequations}
Observe that the velocity autocorrelation times of
$\BV\cN(t/\varepsilon)$ and $\BV\cN_s(t/\varepsilon)$ scale as
$O(\varepsilon)$, which necessitates multiplying both
$\tau_\varepsilon$ and $\tau_{s,\varepsilon}$ by $\varepsilon$
above. Taking into account~\eqref{eq:F_rescale}, for the corresponding
Kolmogorov equation we have
\begin{multline}
\varepsilon\parderiv Ft+\BV V\cdot\nabla_{\BV X}F+\nabla_{\BV V}
\cdot(\BV AF) -\varepsilon\nabla_{\BV X}\cdot\left((\tau_\varepsilon
-\tau_{s, \varepsilon})\left(\nabla_{\BV X}(\theta_\varepsilon F)-\BV{
  \bar A}_\varepsilon F\right)\right)-\\-\nabla_{\BV Y}\cdot\left((
\tau_\varepsilon-\tau_{s,\varepsilon})\left(\nabla_{\BV Y}(
\theta_\varepsilon F)-\BV{\bar A}_\varepsilon F\right) \right)=\frac
1\varepsilon\cL_s^*F.
\end{multline}
This yields, in the corresponding orders of $\varepsilon$,
\begin{subequations}
\begin{equation}
\cL_s^*F_0=0,
\end{equation}
\begin{equation}
\cL_s^*F_1=\BV V\cdot\nabla_{\BV X}F_0+\nabla_{\BV V}\cdot(\BV AF_0)-
\nabla_{\BV Y}\cdot\left(( \tau_0-\tau_{s,0})\left(\nabla_{\BV Y}
(\theta_0 F_0)-\BV{\bar A}_0 F_0\right)\right),
\end{equation}
\begin{multline}
\parderiv{F_0}t+\BV V\cdot\nabla_{\BV X}F_1+\nabla_{\BV V}\cdot (\BV
AF_1)-\nabla_{\BV X}\cdot\left((\tau_0- \tau_{s,0})\left(\nabla_{\BV
  X}(\theta_0 F_0)-\BV{\bar A}_0 F_0\right) \right)-\\-\nabla_{\BV
  Y}\cdot\left((\tau_0-\tau_{s,0})\left( \nabla_{\BV Y}(\theta_0
F_1)-\BV{\bar A}_0 F_1\right)\right) =\cL_s^*F_2.
\end{multline}
\end{subequations}
This leads to $F_0$ having the same meaning as before (a uniform
invariant measure in $\BV Y$ and on the constant momentum/energy
sphere of $\BV V$), which means that the $\BV Y$-derivative of $F_0$
(and, therefore, $\theta_0$ and $\tau_0-\tau_{s,0}$) is zero, which
sets the second identity above to
\begin{equation}
\cL_s^*F_1=\BV V\cdot\nabla_{\BV X}F_0+\nabla_{\BV V}\cdot(\BV A F_0).
\end{equation}
As before, integrating the third equation in $\dif\BV Y\dif\BV V$
yields
\begin{equation}
\label{eq:barF0a}
\parderiv{\bar F_0}t+\int\BV V\cdot\nabla_{\BV X}F_1\,\dif\BV Y\dif
\BV V-\nabla_{\BV X}\cdot\left((\tau_0-\tau_{s,0})\left(\nabla_{\BV X}
(\theta_0\bar F_0)-\BV{\bar A}_0\bar F_0\right)\right)=0.
\end{equation}
Comparing the above equation with what was derived in the previous
section, it becomes clear that, under the same assumptions, the
integral above is given by
\begin{equation}
\label{eq:VF1a}
\int\BV V\cdot\nabla_{\BV X} F_1\,\dif\BV Y\dif\BV V=-\nabla_{\BV X}
\left(\tau_{s,0}\left(\nabla_{\BV X}(\bar F_0\theta_0)-\BV{\bar A}_0
\bar F_0\right)\right).
\end{equation}
Substituting~\eqref{eq:VF1a} into~\eqref{eq:barF0a}, we arrive
at
\begin{equation}
\parderiv{\bar F_0}t=\nabla_{\BV X}\cdot\left(\tau_0\left(\nabla_{\BV
  X}\left(\theta_0\bar F_0\right)-\BV{\bar A}_0\bar F_0\right)\right),
\end{equation}
where we remember that the non-lagged long-term statistics of $\BV\cN$
and $\BV\cN_s$ are identical due to their identical invariant
measures.

\section{The pairwise random interaction operator and the Boltzmann
equation}
\label{sec:boltzmann}

In~\cite{Abr13}, we found that the Boltzmann equation can be obtained
directly from a pairwise random interaction operator which preserves
the momentum and energy of interactions of each pair of molecules.
Here we briefly recall its derivation and show how the Boltzmann
equation can be obtained, together with the corresponding fluid
dynamics equations. Note that the molecules of a real gas do not
interact in this manner. The reason why we consider such an operator
is that for a realistic interaction (i.e. via the Lennard-Jones
potential~\cite{Len}) the subsequent formal derivation of the
Boltzmann equation is not readily available, and thus we have to
resort to a simplification.

Let the simplified molecular interaction process
in~\eqref{eq:dyn_sys_s_slow} with $\varepsilon=0$ be given by the
velocity random jump generator $\cL_s$ of the following form:
\begin{equation}
\label{eq:gen_random_jump}
\cL_s\psi=\BV V\cdot\nabla_{\BV X}\psi+\int G(\BV X',\BV V'|\BV X,\BV
V)\big(\psi(\BV X',\BV V')-\psi(\BV X,\BV V)\big)\dif\BV X'\dif\BV V',
\end{equation}
where the form of the conditional probability $G$ for the random jump
process $\BV\cN_s$ is to be specified below. This simplified process
is known as the L\'evy-type Feller process~\cite{App}, and the form of
its generator above is established by Courr\`ege's theorem~\cite{Cou}.
Despite its simplicity, the random jump generator
in~\eqref{eq:gen_random_jump} can statistically describe a wide
variety of molecular deflections via an appropriate choice of the
conditional probability $G$.

In what follows, $G$ is chosen to be such that the jumps, generated by
it, are reversible (that is, $G(\BV X',\BV V'|\BV X,\BV V)=G(\BV X,\BV
V|\BV X',\BV V')$).  Thus, the approximate system
in~\eqref{eq:dyn_sys_approx} together with the generator
in~\eqref{eq:gen_random_jump} yields the following Kolmogorov
equation:
\begin{multline}
\label{eq:kolm_random_jump}
\parderiv Ft+\BV V\cdot\nabla_{\BV X}F+\nabla_{\BV V}\cdot(\BV A F)=
\nabla_{\BV X}\cdot\left((\tau-\tau_s) \left(\nabla_{\BV X}(\theta F)
-\BV{\bar A} F \right)\right)+\\+\int G(\BV X',\BV V'|\BV X,\BV V)
\big(F(\BV X',\BV V')-F(\BV X,\BV V)\big)\dif\BV X'\dif\BV V'.
\end{multline}
Next, following~\cite{Abr13}, we specify $G$ in the form of pairwise
random interactions of all particles, under the condition that the
coordinates (both translational and rotational) of all molecules do
not change during the jump:
\begin{multline}
G(\BV X',\BV V'|\BV X,\BV V)=\frac 6{\pi\sigma^3V_{rot}}\delta(\BV X'
-\BV X)\cdot\\\cdot\sum_{\myatop{i=1\ldots K-1}{j=i+1\ldots K}}\bigg(
g(\BV v_i',\BV v_j'|\BV v_i,\BV v_j)\chi_\sigma\left(\|\BV x_i^3-\BV
x_j^3\|\right)\prod_{\myatop{k\neq i}{k\neq j}}\delta(\BV v_k'-\BV
v_k) \bigg).
\end{multline}
Above, $\BV x_i^3$ denotes the 3-dimensional translational coordinate
vector of $i$-th molecule, and $\chi_\sigma$ is the following
indicator function:
\begin{equation}
\chi_\sigma(x)=\left\{\begin{array}{l} 1,\qquad 0\leq x\leq\sigma,
\\ 0,\qquad x>\sigma.
\end{array}\right.
\end{equation}
As we can see, each $i,j$-pair of particles is permitted to interact
via a two-particle conditional jump measure $g$ (which is to be
prescribed below) whenever the 3-dimensional translational distance
between them is within $\sigma$, and when they do, all other particles
(enumerated via $k$) retain their original velocities. Thus, the
interaction range parameter $\sigma$ can be interpreted as a diameter
of a gas molecule. The reciprocal of the volume of the ball of
diameter $\sigma$ is included in the constant normalization factor of
$G$. The rotational components of $\BV x_i$ are not used at all in the
simplified interaction model, and the reciprocal of the volume
$V_{rot}$ of the space of the rotational coordinates of a single
molecule is also included in the constant normalization factor of
$G$. In addition to that, we will assume that the velocity
decorrelation time $\tau$ of the real gas also does not depend, to any
non-negligible extent, on the rotational components of $\BV X$, and
thus we can restrict ourselves to the solutions $F$
of~\eqref{eq:kolm_random_jump} which do not depend on any rotational
components of $\BV X$. For now, however, we will proceed with the full
$N$-dimensional notations for the coordinate vectors, while keeping in
mind that the rotational components of the coordinate vectors are not
really present (that is, all their spatial derivatives are zero, while
the integrals can be factored out as constants).

At this point, we are ready to formally derive the Boltzmann
equation. First, we assume that $F$ is invariant under an arbitrary
particle re-enumeration (that is, the particles are statistically
indistinguishable). Second, let us denote the marginals
\begin{equation}
f^{(2)}=\int F\,\dif\BV x_3\ldots\dif\BV x_K\dif\BV v_3\ldots\dif\BV
v_K,\qquad f=\int f^{(2)}\dif\BV x_2\dif\BV v_2.
\end{equation}
Then, integrating~\eqref{eq:kolm_random_jump} over all particles but
one, we obtain
\begin{multline}
\parderiv ft+\BV v\cdot\nabla_{\BV x}f+\nabla_{\BV v}\cdot(\BV a f)=
\nabla_{\BV x}\cdot\left((\tau-\tau_s)\left(\nabla_{\BV x}(\theta f)
-\BV{\bar a} f\right)\right)+\\+(K-1)\frac 6{\pi\sigma^3}\int g(\BV
v',\BV w'|\BV v,\BV w)\chi_\sigma\left(\|\BV x^3-\BV y^3\|\right)
\times\\\times \left(f^{(2)} (\BV x,\BV v',\BV y,\BV w')-f^{(2)}(\BV
x,\BV v,\BV y,\BV w)\right)\dif\BV v' \dif\BV w'\dif\BV y^3\dif\BV w,
\end{multline}
where $V_{rot}$ was cancelled due to the integration over the
rotational coordinates. Also, $\BV a$ is the acceleration vector
(which, as we agreed above, contains only translational nonzero
components) acting on a particle at the location $\BV x$ and moving
with the velocity $\BV v$, while $\BV{\bar a}$ is its respective
average over the invariant measure of the fast interaction process.

We now assume that the variations of $F$ are negligible on the scale
of the interaction range $\sigma$, so that the integration over $\BV
y^3$ can be replaced by multiplication by the volume of a
3-dimensional ball of diameter $\sigma$, which simplifies the equation
above to
\begin{multline}
\parderiv ft+\BV v\cdot\nabla_{\BV x}f+\nabla_{\BV v}\cdot(\BV a f)=
\nabla_{\BV x}\cdot\left((\tau-\tau_s)\left(\nabla_{\BV x}(\theta f)
-\BV{\bar a}f\right)\right)+\\+(K-1)\int g(\BV v',\BV w'|\BV v,\BV
w)\left(f^{(2)}(\BV v',\BV w')-f^{(2)}(\BV v,\BV w)\right)\dif\BV
v'\dif\BV w'\dif\BV w.
\end{multline}
At this point, a standard argument is to assume that $f^{(2)}(\BV
v,\BV w)$ approximately factorizes into the product $f(\BV v)f(\BV
w)$,
\begin{equation}
f^{(2)}(\BV v,\BV w)\approx f(\BV v)f(\BV w),
\end{equation}
which yields
\begin{multline}
\parderiv ft+\BV v\cdot\nabla_{\BV x}f+\nabla_{\BV v}\cdot(\BV a f)=
\nabla_{\BV x}\cdot\left((\tau-\tau_s)\left(\nabla_{\BV x}(\theta f)
-\BV{\bar a} f\right)\right)+\\+(K-1)\int g(\BV v',\BV w'|\BV v, \BV
w)\left(f(\BV v')f(\BV w')-f(\BV v)f(\BV w)\right)\dif\BV v'\dif \BV
w'\dif\BV w.
\end{multline}
By construction, the density $f$ is the single-molecule marginal of
the full probability density $F$, and thus is normalized to one.
However, in order to obtain the standard gas dynamics equations, it is
necessary for the full integral of $f$ over the velocities and
coordinates to yield the total mass of the system of molecules. Thus,
denoting by $m$ the mass of a single molecule, we renormalize $f$ by
the total mass $Km$ of the system as
\begin{equation}
f\to K m f.
\end{equation}
In order to replace the renormalized density in the Boltzmann
equation, we need to include the factor $(K m)^{-1}$ in front of the
collision term, since the latter is quadratic in $f$. As a result, we
arrive at the following general form of the Boltzmann equation:
\begin{multline}
\parderiv ft+\BV v\cdot\nabla_{\BV x}f+\nabla_{\BV v}\cdot(\BV a f)=
\nabla_{\BV x}\cdot\left((\tau-\tau_s)\left(\nabla_{\BV x}(\theta f)
-\BV{\bar a} f\right)\right)+\\+\frac{K-1}{Km}\int g(\BV v',\BV w'|\BV
v,\BV w)\left(f(\BV v')f(\BV w')-f(\BV v)f(\BV w)\right)\dif\BV v'
\dif\BV w'\dif\BV w.
\end{multline}
Note that at this stage the momentum and energy conservation laws are
yet to be taken into account, by imposing appropriate constraints on
the conditional jump measure $g$.

In order to prescribe $g$, we need to recall that the interactions
between two particles must preserve their momentum and energy, that
is, the following identities should be in effect:
\begin{equation}
\label{eq:mom_en}
\BV v_i'+\BV v_j'=\BV v_i+\BV v_j,\qquad\|\BV v_i'\|^2+\|\BV
v_j'\|^2=\|\BV v_i\|^2+\|\BV v_j\|^2.
\end{equation}
Let us denote by $\BV n$ the unit vector pointing along the difference
in outgoing velocities $\BV v'$ and $\BV w'$:
\begin{equation}
\BV n=\frac{\BV v'-\BV w'}{\|\BV v'-\BV w'\|}.
\end{equation}
Now, observe that the momentum and energy conservation constraints
in~\eqref{eq:mom_en} imply the following relations between the
incident and outgoing pairs of velocities:
\begin{equation}
\label{eq:vwn}
\BV v'=\frac {\BV v+\BV w}2+\frac{\|\BV v-\BV w\|}2\BV n,\qquad\BV
w'=\frac {\BV v+\BV w}2-\frac{\|\BV v-\BV w\|}2\BV n.
\end{equation}
For realistic particles which move deterministically, the vector $\BV
n$ is obviously not arbitrary and depends on how the particles are
positioned with respect to each other prior to the interaction.
However, in our artificial random jump generator above, the direction
$\BV n$ is a random parameter, since $g$ does not have any information
about the coordinates of the interacting particles. In addition,
observe that under the momentum and energy conservation constraints,
the norm of the difference $\|\BV v-\BV w\|$ is an invariant,
\begin{equation}
\|\BV v'-\BV w'\|=\|\BV v-\BV w\|,
\end{equation}
and thus the integration over $\dif\BV v'\dif\BV w'$ is equivalent to
the integration over $\|\BV v-\BV w\|\dif\BV n$:
\begin{equation}
\label{eq:b}
\frac{K-1}K g(\BV v',\BV w'|\BV v,\BV w)\dif\BV v'\dif\BV w'\sim
\sigma^2 b(\BV v',\BV w'|\BV v,\BV w)\|\BV v-\BV w\|\dif\BV n.
\end{equation}
Above, $\sigma^2$ is the collision cross section, and the interaction
kernel $b$ specifies the random distribution properties of particle
scattering within the limits permitted by the momentum and energy
conservation constraints in~\eqref{eq:mom_en}. Also, we discard the
ratio $(K-1)/K$, since it approaches 1 in the case of many
molecules. As a result, we finally arrive at the diffusive Boltzmann
equation~\cite{Abr13}:
\begin{subequations}
\begin{equation}
\parderiv ft+\BV v\cdot\nabla_{\BV x}f+\nabla_{\BV v}\cdot(\BV a f)=
\coll(f)+\nabla_{\BV x}\cdot\left((\tau-\tau_s)\left(\nabla_{\BV x}
(\theta f)-\BV{\bar a} f\right)\right),
\end{equation}
\begin{equation}
\label{eq:collision}
\coll(f)=\frac{\sigma^2}m\int b(\BV v',\BV w'|\BV v,\BV w)\|\BV v-\BV
w\|\left(f(\BV v')f(\BV w')-f(\BV v)f(\BV w)\right)\dif\BV n\dif\BV w,
\end{equation}
\end{subequations}
where $\BV v'$ and $\BV w'$ are the functions of $\BV v$, $\BV w$ and
$\BV n$, given in~\eqref{eq:vwn}. Note that the collision operator
$\coll(f)$ in general employs the full $N$-dimensional velocity
vectors, with both translational and rotational components (only the
rotational components of the coordinate vectors are discarded in the
presented random collision model). Further simplifications
of~\eqref{eq:collision} can potentially be made for particular
applications (for example, the computation of the theoretical
diffusion coefficient of a random deflection model does not require
the rotational components of the velocity deflection
angle~\cite{HirBirSpo}), however, such direct computations are beyond
the scope of the present work.

There is a variety of choices for $b$~\cite{Cer2,CerIllPul,ChaCow}. In
the simplest case of monatomic molecules ($N=3$) it can be assumed
that particles statistically deflect from each other as if they were
hard spheres of diameter $\sigma$. Then, $b$ depends only on the angle
$\phi$ between the directions of differences in incoming and outgoing
velocities. This case is very well studied in the literature (see, for
example, \cite{Cer2,CerIllPul,Gols,Gra} and references therein). In
more elaborate deflection models~\cite{ChaCow,HirBirSpo,HirBirSpo2},
such as those employing the Lennard-Jones potential~\cite{Len}, the
interaction kernel $b$ also takes into account the relative speed
$\|\BV v-\BV w\|$ of the colliding molecules, in addition to the
deflection angle. Note that these forms of $b$ are automatically
consistent with the reversibility condition which we imposed earlier
on $G$.

At this point, we have to use the single-particle distribution $f$ to
express $\theta$ and $\tau$, rather than the full distribution
$F$. However, it is fairly straightforward under the previous
assumption of statistical identity of the particles. First, let us
denote the density $\rho$, normalized distribution $\tilde f$, and its
moment $\langle\psi\rangle$, for a suitable $\psi(\BV v)$, as
\begin{equation}
\rho = \int f\dif\BV v,\qquad\tilde f=\frac
f\rho,\qquad\langle\psi\rangle=\int\psi\tilde f\dif\BV v,
\end{equation}
respectively. Then, we express the average acceleration $\BV{\bar a}$
and velocity $\BV u$ as
\begin{equation}
\BV{\bar a}=\langle\BV a\rangle,\qquad\BV u=\langle\BV v\rangle,
\end{equation}
where the latter is, by assumption, the same for~\eqref{eq:dyn_sys}
and~\eqref{eq:dyn_sys_s}, since those are presumed to be identical in
the averaging time limit. With this, we express $\theta$ and $\tau$ as
\begin{equation}
\theta=\frac 1N\langle\|\BV v-\BV u\|^2\rangle,\qquad\tau=\frac 1{N
  \theta}\int_0^\infty\langle(\EE\BV v(s)-\BV u)\cdot(\BV v(0)-\BV u)
\rangle\dif s.
\end{equation}
Now let us recall that the conventional $\rho$-weighted diffusion
coefficient $D$ is given by~\cite{Zwa2}
\begin{equation}
\label{eq:D}
D=\frac 1N\int_0^\infty\dif s\int(\EE\BV v(s)-\BV u)\cdot(\BV v(0)-\BV
u) f\dif\BV v=\rho\theta\tau.
\end{equation}
Denoting the pressure $p$ as the product of the density and temperature,
\begin{equation}
p=\rho\theta,
\end{equation}
we can express the decorrelation time $\tau$ as
\begin{equation}
\tau=\frac Dp,\qquad\tau_s=\frac{D_s}p,
\end{equation}
where the pressures of the realistic and simplified systems are
presumed to be identical due to $p$ being the product of non-lagged
statistical quantities $\rho$ and $\theta$.  Also, for convenience, we
follow the notations in~\cite{Abr13} and denote the difference between
$D$ and $D_s$ as a corrective diffusion coefficient $D_\alpha$:
\begin{equation}
\label{eq:D_alpha_2}
D_\alpha=D-D_s.
\end{equation}
In what is to follow, we show that $D_\alpha$ is proportional to the
diffusion coefficient $D$ of the realistic gas interaction process
$\BV\cN$ of~\eqref{eq:dyn_sys}, that is,
\begin{equation}
\label{eq:D_alpha_3}
D_\alpha=\alpha D,\qquad \alpha=\frac{D-D_s}D>0.
\end{equation}
In~\cite{Abr13} and~\cite{AbrOtt}, the diffusion coefficient
$D_\alpha$ (and, therefore, the proportionality coefficient $\alpha$)
were set empirically via numerical simulations. In what is to follow,
we compute the formulas for the estimates of $\alpha$ and $D_\alpha$
based on the known properties of the system. Finally, the Boltzmann
equation can be written via $D_\alpha$ as
\begin{equation}
\label{eq:boltzmann}
\parderiv ft+\BV v\cdot\nabla_{\BV x}f+\nabla_{\BV v}\cdot(\BV a f)=
\coll(f)+\nabla_{\BV x}\cdot\left(D_\alpha\nabla_{\BV x}\tilde f
\right) + \nabla_{\BV x}\cdot\left(\frac{D_\alpha}p (\nabla_{\BV x}p
-\rho\langle\BV a\rangle)\tilde f\right).
\end{equation}
Observe that the expression $\nabla_{\BV x}p-\rho\langle\BV a\rangle$
in the right-hand side above is zero when the hydrostatic balance is
achieved.

In what follows, we recognize the fact that the previously estimated
values of the diffusion coefficient for different gases, either in the
case of the relatively simple hard sphere collision
model~\cite{Cer2,CerIllPul,Gols,Gra}, or in the case of the more
sophisticated Lennard-Jones random deflection
model~\cite{ChaCow,HirBirSpo,HirBirSpo2}, are all based on the general
collision operator of the form~\eqref{eq:collision}, which in the
presented theory corresponds to the simplified random gas process
$\BV\cN_s$. Because of this fact, there is no need for us to carry out
any additional computations to estimate the corrective diffusion
coefficient $D_\alpha$ for gases which were already studied in the
aforementioned works; the theoretical values of the diffusion
coefficient $D_s$ for a given simplified random deflection model can
be found in \cite{ChaCow,HirBirSpo,HirBirSpo2}, whereas the values of
the diffusion coefficient $D$ of the corresponding real gas can be
looked up in the works where the direct laboratory measurements have
been completed, such as~\cite{Hut,Winn}.

\section{The Euler equations}
\label{sec:Euler}

For a suitable $\psi(\BV v)$, let us denote its collision moment as
\begin{equation}
\langle\psi\rangle_{\coll(f)}=\int\psi\,\coll(f)\dif\BV v.
\end{equation}
Then, integrating the Boltzmann equation in~\eqref{eq:boltzmann}
against $\psi\dif\BV v$, we obtain
\begin{multline}
\parderiv{(\rho\langle\psi\rangle)}t+\nabla_{\BV x}\cdot(\rho\langle
\psi\BV v\rangle)=\rho\langle(\BV a\cdot\nabla_{\BV v})\psi\rangle
+\langle\psi\rangle_{\coll(f)}+\\+\nabla_{\BV x}\cdot\left(D_\alpha
\nabla_{\BV x}\langle\psi\rangle\right)+\nabla_{\BV x}\cdot \left(
\frac{D_\alpha}p(\nabla_{\BV x}p-\rho\langle\BV a\rangle)\langle\psi
\rangle\right),
\end{multline}
where we integrated the term with the external acceleration $\BV a$ by
parts. In what follows, we will for simplicity assume that $\BV a$
does not depend on the particle velocity $\BV v$, that is,
\begin{equation}
\BV a=\BV a(\BV x),\qquad\langle\BV a\rangle=\BV a,\qquad\langle(\BV
a\cdot \nabla_{\BV v})\psi\rangle=\BV a\cdot\langle\nabla_{\BV v}\psi
\rangle.
\end{equation}
If necessary, what is presented below can be recalculated for a $\BV
v$-dependent $\BV a$.

Observe that, due to mass, momentum and energy conservation, the
following collision moments are zero:
\begin{equation}
\langle 1\rangle_{\coll(f)}=0,\qquad\langle\BV v\rangle_{\coll(f)}=\BV 0,
\qquad\langle\|\BV v\|^2\rangle_{\coll(f)}=0.
\end{equation}
Denoting the energy
\begin{equation}
\label{eq:E_theta}
E=\frac 12\langle\|\BV v\|^2\rangle=\frac 12(\|\BV u\|^2+N\theta),
\end{equation}
we obtain, correspondingly, the mass, momentum and energy transport
equations:
\begin{subequations}
\label{eq:rho_u_E}
\begin{equation}
\parderiv\rho t+\nabla\cdot(\rho\BV u)=\nabla\cdot\left(\frac{
  D_\alpha} p(\nabla p-\rho\BV a)\right),
\end{equation}
\begin{equation}
\parderiv{(\rho\BV u)}t+\nabla\cdot(\rho\langle\BV v\BV v^T\rangle)=
\rho\BV a+\nabla\cdot\left(D_\alpha\nabla\BV u\right)+\nabla\cdot
\left(\frac{D_\alpha}p(\nabla p-\rho\BV a)\BV u^T\right),
\end{equation}
\begin{equation}
\parderiv{(\rho E)}t+\frac 12\nabla\cdot(\rho\langle\|\BV v\|^2\BV v
\rangle)=\rho\BV u\cdot\BV a+\nabla\cdot\left(D_\alpha\nabla E\right)
+\nabla\cdot\left(\frac{D_\alpha}p(\nabla p-\rho\BV a)E\right).
\end{equation}
\end{subequations}
The higher-order moments in the advection terms can be expressed as
\begin{equation}
\label{eq:advection_moments}
\langle\BV v\BV v^T\rangle=\BV u\BV u^T+\BM T,\qquad\frac
12\langle\|\BV v\|^2\BV v\rangle=E\BV u+\BM T\BV u+\BV q,
\end{equation}
where the temperature tensor $\BM T$ and heat flux $\BV q$ are given
by the velocity-centered moments
\begin{equation}
\label{eq:T_q}
\BM T=\langle(\BV v-\BV u)(\BV v-\BV u)^T\rangle,\qquad\BV q=\frac
12\langle\|\BV v-\BV u\|^2(\BV v-\BV u)\rangle.
\end{equation}
In the conventional Euler equations~\cite{Bat,Gols}, $\BM T$ and $\BV
q$ are set to
\begin{equation}
\label{eq:T_q_Euler}
\BM T_E=\theta\BM I,\qquad\BV q_E=\BV 0,
\end{equation}
and thus the equations in~\eqref{eq:rho_u_E} become fully closed with
respect to $\rho$, $\BV u$ and $E$. In the current set-up, however,
the identities in~\eqref{eq:T_q_Euler} require a suitable
justification, which, in turn, should be based on the properties of
the simplified interaction process $\BV\cN_s$.

Let us, for convenience, denote the stress $\BM S$ as the difference
between $\BM T$ and $\theta\BM I$:
\begin{equation}
\label{eq:S}
\BM S=\BM T-\theta\BM I,\qquad \BM S_E=\BM 0,
\end{equation}
where the second identity follows immediately
from~\eqref{eq:T_q_Euler}.  Observe that for $\varepsilon=0$ the
solution of~\eqref{eq:dyn_sys_s_slow} should eventually settle at the
uniformly distributed dynamics on the constant momentum-energy sphere
due to strong mixing. Clearly, when the dimension of the sphere is
large enough (so that the single-particle marginal distributions are
Gaussian~\cite{Abr13}), both the stress $\BM S$ in~\eqref{eq:S} and
heat flux $\BV q$ in~\eqref{eq:T_q} become zero. At the same time, the
rate at which $\BM S$ and $\BV q$ approach zero is determined by the
rate of interactions in the simplified process $\BV\cN_s$ (the more
interactions occur per unit of time, the faster $\BM S$ and $\BV q$
approach zero). Thus, a consistent way in the present framework to set
$\BM S$ and $\BV q$ of the approximate fast interaction process
$\BV\cN_s$ in~\eqref{eq:dyn_sys_approx} much closer to zero than their
realistic counterparts of $\BV\cN$ is to ``accelerate'' $\BV\cN_s$ by
the time rescaling $\BV\cN_s(t)\to\BV\cN_s(t/\delta)$ for some
constant $0<\delta\ll 1$. In this case, by adjusting $\delta$, the
stress $\BM S$ and heat flux $\BV q$ of~\eqref{eq:dyn_sys_approx} can
be damped to zero as fast as necessary, so that one would be able to
discard them in comparison with the realistic stress and heat flux
of~\eqref{eq:dyn_sys}.

However, the same time rescaling of $\BV\cN_s$ also scales its
corresponding diffusion coefficient $D_s$ by the same constant
parameter $\delta$ as above, as $D_s$ is given by the correlation time
integral~\eqref{eq:D}. In this situation, the diffusion coefficient
$D_s$ of the simplified process $\BV\cN_s$ should also become
negligible in comparison with the realistic diffusion coefficient
$D$. Therefore, the following identities follow
from~\eqref{eq:D_alpha_2} and~\eqref{eq:D_alpha_3} for the Euler
equations:
\begin{equation}
\alpha=\frac{D-D_s}D=1,\qquad\text{or}\qquad D_\alpha=D.
\end{equation}
Thus, the Euler equations assume the form
\begin{subequations}
\label{eq:Euler}
\begin{equation}
\parderiv\rho t+\nabla\cdot(\rho\BV u)=\nabla\cdot\left(\frac Dp
(\nabla p-\rho\BV a)\right),
\end{equation}
\begin{equation}
\parderiv{(\rho\BV u)}t+\nabla\cdot(\rho(\BV u\BV u^T+\theta\BM I))=
\rho\BV a+\nabla\cdot\left(D\nabla\BV u\right)+\nabla\cdot\left(\frac
Dp(\nabla p-\rho\BV a)\BV u^T\right),
\end{equation}
\begin{equation}
\parderiv{(\rho E)}t+\nabla\cdot(\rho(E+\theta)\BV u)=\rho\BV u\cdot
\BV a+\nabla\cdot\left(D\nabla E\right)+\nabla\cdot\left(\frac Dp
(\nabla p-\rho\BV a)E\right).
\end{equation}
\end{subequations}
Formally, the Euler equations above are derived for the full
$N$-dimensional dynamics, which involve translational and rotational
(if any) components of the average velocity vector $\BV u$. However,
observe that the all components of $\BV u$ beyond translational can be
discarded from the equations. Indeed, the rotational components of
$\BV u$ never enter the advection terms due to the corresponding zero
spatial derivatives, as $f$ does not depend on them. Also, they do not
enter the external acceleration term because the corresponding
components of $\BV a$ are also zero. Thus, the Euler equations above
in~\eqref{eq:Euler} remain valid if the average velocity vector $\BV
u$, the external acceleration vector $\BV a$, and the spatial
differentiation operator $\nabla$ are all truncated to the first 3
translational components.

Observe that the additional diffusion terms, which are present in the
right-hand side of~\eqref{eq:Euler} (and absent from the conventional
Euler equations~\cite{Bat,Gols}) are nonvanishing, since it is
impossible for the diffusion coefficient $D$ of a realistic gas to be
zero as long as there is a free flight between consecutive molecular
collisions (which is a requirement for a dilute
gas~\cite{Cer2,CerIllPul,Gra}). Thus, the conventional Euler
equations~\cite{Bat,Gols} do not appear to be a valid approximation
for a realistic gas on a time scale longer than $O(1)$, and the
corrected equations in~\eqref{eq:Euler} should be used instead. In
Tables~\ref{tab:NS_gases} and~\ref{tab:NS_gases_2} we show the
measured diffusion coefficients for some gases.

\section{The Navier-Stokes equations}
\label{sec:Navier-Stokes}

Unlike the Euler equations above, the Navier-Stokes
equations~\cite{Bat,Gols} appear when the moment transport equations
in~\eqref{eq:rho_u_E} contain appropriately parameterized expressions
for the stress $\BM S$ and heat flux $\BV q$, given by the Newton and
Fourier laws:
\begin{subequations}
\label{eq:S_q_NS}
\begin{equation}
\BM S_{NS}=\frac\mu\rho\left(\frac 2N(\nabla\cdot\BV u)\BM I-\nabla\BV u-(
\nabla\BV u)^T\right),
\end{equation}
\begin{equation}
\BV q_{NS}=-\frac{N+2}{2\,\Pran}\frac\mu\rho\nabla\theta.
\end{equation}
\end{subequations}
Just as the Euler equations, the Navier-Stokes equations formally
contain the velocity moments for all degrees of freedom, however, they
can be restricted to the translational components of $\BV u$, $\BM S$
and $\BV q$ using the same reasoning as for the Euler equations.

Above in~\eqref{eq:S_q_NS}, the parameter $\Pran$ is the Prandtl
number, and the coefficient $\mu$ is the viscosity, given, in the
absence of a long-range interaction potential and under the assumption
of statistical equivalency of the velocity components, by the
following correlation integral \cite{Gre,Kub1,Zwa2}:
\begin{equation}
\label{eq:mu}
\mu=\frac 1\theta\int_0^\infty\dif s\int\EE\big[(v(s)-u)(v'(s)-u')]
(v(0)-u)(v'(0)-u') f\dif\BV v.
\end{equation}
Here, $v$ and $v'$ are two different, but otherwise arbitrary,
molecular velocity components, while $u$ and $u'$ are their respective
averages. It is clear that a time-rescaling of the simplified
interaction process $\BV\cN_s$ adjusts both its diffusion coefficient
$D_s$ and its viscosity $\mu_s$ by the same factor (since both are
time autocorrelation functions). Recall that the Schmidt number
$\Schm$ of a gas is given by the quotient of the viscosity $\mu$ and
the mass diffusion coefficient $D$:
\begin{equation}
\label{eq:Schmidt_number}
\Schm=\frac\mu D.
\end{equation}
Clearly, the Schmidt number $\Schm_s$ of the simplified interaction
process $\BV\cN_s$ is invariant under the time rescaling of the
latter.

In what follows, we construct the Navier-Stokes equations by choosing
the appropriate time rescaling for $\BV\cN_s$ so that the viscosity of
the simplified interaction process $\BV\cN_s$ matches that of
$\BV\cN$. The corrective diffusion coefficient $D_\alpha$ is then
chosen to simultaneously equalize the diffusion rates of the realistic
and approximate gas systems. More precisely, let us assume that the
activity of the collision kernel $b$ in~\eqref{eq:b} is normalized so
that $\mu_s=\mu$, where the latter is the experimentally measured
autocorrelation function~\eqref{eq:mu} for a given real gas. Then, we
can compute the corresponding diffusion coefficients, $D_s$ and $D$,
by expressing them via the corresponding Schmidt numbers:
\begin{equation}
D=\frac\mu\Schm,\qquad D_s=\frac\mu{\Schm_s}=\frac\Schm {\Schm_s}D.
\end{equation}
Subsequently, from~\eqref{eq:D_alpha_2} and~\eqref{eq:D_alpha_3} we
can express $\alpha$ and $D_\alpha$ as
\begin{equation}
\label{eq:D_alpha}
\alpha=\frac{D-D_s}D=1-\frac\Schm{\Schm_s},\qquad D_\alpha=\left(
1-\frac\Schm{ \Schm_s}\right)D=\left(\frac 1\Schm-\frac 1{\Schm_s}
\right) \mu.
\end{equation}
As we can see, the corrective diffusion coefficient $D_\alpha$ is
expressed via the viscosity of the real gas, and the Schmidt numbers
of the real and simplified gas processes. The Schmidt number of a real
gas can be determined as a quotient of the measured viscosity and mass
diffusion coefficient, while the one of a simplified process
$\BV\cN_s$ can, in some cases, be evaluated from the properties of
$\BV\cN_s$. In particular, the Schmidt number of a random hard sphere
deflection process has been computed explicitly~\cite{ChaCow,Hey,Dym}
under the same assumptions as were used above for the derivation of
the Boltzmann equation in~\eqref{eq:boltzmann} (that is, statistical
independence and identical distribution of molecules):
\begin{equation}
\label{eq:Sc_HS}
\Schm_{HS}=\frac 56.
\end{equation}
The diffusion properties and the corresponding Schmidt numbers of real
gases are also an extensively studied subject, and the relevant data
is readily available~\cite{KayLab,Hut}. In Table~\ref{tab:NS_gases} we
show the measured viscosity, Schmidt number, diffusion coefficient,
and the corresponding coefficients $\alpha_{HS}$ and $D_{\alpha,HS}$
(the latter two computed from~\eqref{eq:D_alpha} and~\eqref{eq:Sc_HS})
for neon, argon, krypton and xenon.
\begin{table}
\begin{tabular}{|c||c|c|c|c|c|}
\hline
Gas & $\mu$ & $\Schm$ & $D$ & $\alpha_{HS}$ & $D_{\alpha,HS}$ \\
\hline\hline
Ne & $3.1\cdot 10^{-5}$ & $0.787$ & $3.9\cdot 10^{-5}$ & $5.51\cdot 10^{-2}$ & $2.2\cdot 10^{-6}$ \\
Ar & $2.2\cdot 10^{-5}$ & $0.763$ & $2.9\cdot 10^{-5}$ & $8.4\cdot 10^{-2}$ & $2.4\cdot 10^{-6}$ \\
Kr & $2.5\cdot 10^{-5}$ & $0.769$ & $3.3\cdot 10^{-5}$ & $7.69\cdot 10^{-2}$ &$2.5\cdot 10^{-6}$ \\
Xe & $2.3\cdot 10^{-5}$ & $0.806$ & $2.9\cdot 10^{-5}$ & $3.23\cdot 10^{-2}$ & $9.2\cdot 10^{-7}$ \\
\hline
\end{tabular}
\vspace{1EX}
\caption{The diffusive properties of some monatomic gases at
  20$^\circ$ C (except for argon at 22$^\circ$ C) based on the hard
  sphere collision model. The viscosity values $\mu$ are taken
  from~\cite{KayLab}, while the measured Schmidt numbers $\Schm$ are
  taken from~\cite{Hut}.  The corresponding values $D$ of the
  diffusion coefficient are computed from the definition of the
  Schmidt number in~\eqref{eq:Schmidt_number}. The units of the
  viscosity and the mass diffusion coefficient are [kg (m s)$^{-1}$].
  The coefficient $\alpha_{HS}$ and the corrective diffusion
  $D_{\alpha,HS}$ are computed from~\eqref{eq:D_alpha}
  and~\eqref{eq:Sc_HS}.}
\label{tab:NS_gases}
\end{table}

In the Lennard-Jones random deflection model (which was adapted for
monatomic as well as polyatomic
gases~\cite{HirBirSpo,HirBirSpo2,Winn}), the collision parameters are
also chosen so that the resulting model viscosity coefficients match
the measured values, and thus we can estimate the corresponding
coefficients $\alpha_{LJ}$ and $D_{\alpha,LJ}$ from the difference
between the measured diffusion coefficient $D$ and its Lennard-Jones
estimate $D_{LJ}$ via~\eqref{eq:D_alpha}. In this case, both the
theoretical estimated~\cite{HirBirSpo,HirBirSpo2} and
measured~\cite{Winn} diffusion coefficients are provided directly, and
thus there is no need to express them via the viscosity and Schmidt
numbers. We show the corresponding computed coefficients
$\alpha_{LJ}$ and $D_{\alpha,LJ}$ in Table~\ref{tab:NS_gases_2} for
neon, argon, nitrogen, oxygen, carbon dioxide and methane. Observe
that the Lennard-Jones coefficients $\alpha_{LJ}$ and $D_{\alpha,LJ}$
in Table~\ref{tab:NS_gases_2} for neon and argon are much smaller than
their hard sphere counterparts $\alpha_{HS}$ and $D_{\alpha,HS}$ in
Table~\ref{tab:NS_gases}. This is likely due to the fact that the
Lennard-Jones random deflection model for the monatomic gases is more
accurate than the hard sphere random collision model, and subsequently
necessitates a smaller diffusion correction.
\begin{table}
\begin{tabular}{|c||c|c|c|c|}
\hline
Gas & $D$ & $D_{LJ}/D$ & $\alpha_{LJ}$ & $D_{\alpha,LJ}$ \\
\hline\hline
Ne & $4.3\cdot 10^{-5}$ & $0.977$ & $2.33\cdot 10^{-2}$ & $9.9\cdot 10^{-7}$ \\
Ar & $2.9\cdot 10^{-5}$ & $0.989$ & $1.11\cdot 10^{-2}$ & $3.3\cdot 10^{-7}$ \\
N$_2$ & $2.4\cdot 10^{-5}$ & $0.958$ & $4.25\cdot 10^{-2}$ & $10^{-6}$ \\
O$_2$ & $3\cdot 10^{-5}$ & $0.888$ & $0.112$ & $3.4\cdot 10^{-6}$ \\
CO$_2$ & $2\cdot 10^{-5}$ & $0.965$ & $3.54\cdot 10^{-2}$ & $7.2\cdot 10^{-7}$ \\
CH$_4$ & $1.6\cdot 10^{-5}$ & $0.895$ & $0.104$ & $1.6\cdot 10^{-6}$ \\
\hline
\end{tabular}
\vspace{1EX}
\caption{The diffusive properties of some monatomic and polyatomic
  gases at 25$^\circ$ C (except for argon at 22$^\circ$ C) based on
  the Lennard-Jones potential~\cite{Len} model. The data is taken
  from~\cite{Winn,HirBirSpo,HirBirSpo2}. The units of the the mass
  diffusion coefficient are [kg (m s)$^{-1}$].  The coefficient
  $\alpha_{LJ}$ and the corrective diffusion $D_{\alpha,LJ}$ are
  computed from~\eqref{eq:D_alpha}.}
\label{tab:NS_gases_2}
\end{table}

Previously in~\cite{Abr13,AbrOtt}, we used the following empirically
chosen values of $D_\alpha$ for argon at normal conditions: $4\cdot
10^{-6}$ in~\cite{Abr13} (chosen to match the temperature profile of
the DSMC~\cite{Bird,ScaRooWhiDarRee} Couette flow for the
Navier-Stokes equations), and $10^{-6}$ in~\cite{AbrOtt} (chosen to
match the velocity profile of the DSMC Poiseuille flow for the
Navier-Stokes equations). Observe that, even though the Navier-Stokes
equations were matched to a DSMC computation rather than the actual
deterministic gas, those empirical values of $D_\alpha$ were not too
far from the theoretical estimates of $2.4\cdot 10^{-6}$ (hard sphere)
and $3.3\cdot 10^{-7}$ (Lennard-Jones) for argon that we show in
Tables~\ref{tab:NS_gases} and~\ref{tab:NS_gases_2}. For nitrogen, the
Lennard-Jones value of $10^{-6}$ in Table~\ref{tab:NS_gases_2}
coincidentally matches our empirical choice in~\cite{AbrOtt}.

\section{The Grad equations}
\label{sec:Grad}

The matching viscosity approximation in~\eqref{eq:D_alpha}, used above
for the Navier-Stokes equations, can be naturally extended onto the
Grad~\cite{Gra,Mal} and regularized Grad~\cite{Abr13,StruTor}
equations. The transport equations for the mass, momentum and energy
in~\eqref{eq:rho_u_E} are complemented by the additional transport
equations for the stress and heat flux, which are derived by
integrating the Boltzmann equation in~\eqref{eq:boltzmann} against the
second and third velocity moments, and subtracting the transport
equations for lower-order moments in appropriate combinations:
\begin{subequations}
\label{eq:S_q}
\begin{multline}
\parderiv{(\rho\BM S)}t+\nabla\cdot(\rho(\BV u\otimes\BM S+\BM Q))+
\left(\BM P+\BM P^T-\frac 2N\trace(\BM P)\BM I\right)=\\=-\frac{
  \rho p}\mu\BM S+\nabla\cdot\left(D_\alpha\nabla\otimes\BM S\right)
+\nabla\cdot\left(\frac{D_\alpha}p(\nabla p-\rho\BV a)\otimes\BM S
\right),
\end{multline}
\begin{multline}
\parderiv{(\rho\BV q)}t+\nabla\cdot(\rho\BV u\BV q^T)+\nabla\cdot(\BM
P_2\BM T_2)-\BM T_2\nabla\cdot\BM P_2-\nabla\cdot(\rho\BM S^2)+\rho(
\nabla\BV u)^T\BV q+\\+\frac 2{N+2}\rho\left[\nabla\BV u+(\nabla\BV
  u)^T+(\nabla\cdot\BV u)\BM I\right]\BV q+\rho\BM Q:(\nabla\BV u)
+\nabla\cdot(\rho\BM R)=\\=-\Pran\frac{\rho p}\mu\BV q+\nabla\cdot
\left(D_\alpha\nabla\BV q\right)+\nabla\cdot\left(\frac{D_\alpha}p
(\nabla p-\rho\BV a)\BV q^T)\right),
\end{multline}
\end{subequations}
where
\begin{subequations}
\begin{equation}
\BM P=\left(\rho\BM T-D_\alpha(\nabla\BV u)^T\right)\nabla\BV u+\frac
2{N+2}\nabla(\rho\BV q),
\end{equation}
\begin{equation}
\BM T_2=\BM T+\frac N2\theta\BM I, \qquad\BM P_2=\rho\BM T-2D_\alpha
\nabla\BV u.
\end{equation}
\end{subequations}
Observe that the equations in~\eqref{eq:rho_u_E} and~\eqref{eq:S_q}
are not closed with respect to the matrix $\BM R$ and the 3-rank
tensor $\BM Q$ above. Above, both the rank-3 tensor $\BM Q$ and the matrix $\BM R$ equal
zero for the Grad equations \cite{Gra,Mal}, while for the regularized
Grad equations~\cite{StruTor,Abr13} these are set to
\begin{subequations}
\label{eq:Reg_Grad}
\begin{equation}
\BM Q_{reg}=\widetilde{\!\BM Q}+\widetilde{\!\BM Q}^T+\widetilde{\!\BM
  Q}^{TT},
\end{equation}
\begin{equation}
\BM R_{reg}=\widetilde{\!\BM R}+\widetilde{\!\BM R}^T+\left(\widetilde
R-\frac 2N\trace(\widetilde{\!\BM R})\right)\BM I,
\end{equation}
\end{subequations}
where the notations $\widetilde{\!\BM Q}$, $\widetilde R$ and
$\widetilde{\!\BM R}$ read
\begin{subequations}
\label{eq:Reg_Grad_2}
\begin{multline}
\rho\,\widetilde{\!\BM Q}=-\frac\mu{\Pran^{\widetilde{\BM Q}}}\bigg[
  \nabla \BM S-\frac 2{N+2}\BM I\otimes\nabla\cdot\BM S-\frac 1p
  \left(\BM S \otimes\nabla\cdot(\rho\BM S)-\frac 2{N+2}\BM I\otimes
  \BM S\,\nabla\cdot(\rho\BM S)\right)+\\+\frac 2{(N+2)\theta}\bigg(
  \BV q\otimes\left(\nabla\BV u+(\nabla\BV u)^T\right)-\frac 2{N+2}\BM
  I\otimes\left(\nabla\BV u+ (\nabla\BV u)^T+(\nabla\cdot\BV u)\BM I
  \right)\BV q\bigg)\bigg],
\end{multline}
\begin{equation}
\rho\widetilde R=-\frac{2\mu}{\Pran^{\widetilde R}}\left[\frac{N+2}N
  \BV q\cdot\frac{\nabla\theta}\theta+\frac 2N\left(\nabla\cdot\BV q+
  \BM S:(\nabla\BV u)-\frac 1p\BV q^T\nabla\cdot(\rho\BM S)\right)
  \right],
\end{equation}
\begin{multline}
\rho\,\widetilde{\!\BM R}=-\frac\mu{\Pran^{\widetilde{\BM R}}}\bigg[
  \BM S\left(\nabla\BV u+(\nabla\BV u)^T\right)+\frac{N+4}{N+2}
  \left(\frac{\nabla(\theta\BV q)}\theta-\frac 1p\nabla\cdot( \rho\BM
  S)\BV q^T\right)-\\-\left(\frac 2N\nabla\cdot\BV u+\frac{N+4}{2N
    \theta}\left(\frac 1\rho\nabla\cdot(\rho\BV q)+\BM S:(\nabla\BV u)
  \right)\right)\BM S\bigg].
\end{multline}
\end{subequations}
Above, the constants $\Pran^{\widetilde{\BM Q}}$, $\Pran^{\widetilde
  R}$ and $\Pran^{\widetilde{\BM R}}$ are the third- and fourth-order
moment Prandtl numbers, which, for a hard sphere collision
approximation in the case of $N=3$ \cite{StruTor}, are equal to
\begin{equation}
\Pran^{\widetilde{\BM Q}}_{HS}=\frac 32,\qquad\Pran^{\widetilde
  R}_{HS}=\frac 23,\qquad \Pran^{\widetilde{\BM R}}_{HS}=\frac 76.
\end{equation}
As it seems, there are currently no estimates of
$\Pran^{\widetilde{\BM Q}}$, $\Pran^{\widetilde R}$ and
$\Pran^{\widetilde{\BM R}}$ for any other collision models, and, in
particular, we used the hard sphere values above for nitrogen
in~\cite{Abr13,AbrOtt}, which still provided reasonable accuracy of
the results of numerical simulations. The value of the corrective
diffusion coefficient $D_\alpha$ remains the same as before for the
Navier-Stokes equations in~\eqref{eq:D_alpha}, and thus the
approximations from Tables~\ref{tab:NS_gases} and~\ref{tab:NS_gases_2}
can be used for the listed gases.

\section{Summary}
\label{sec:summary}

Above we proposed a theory for the mass diffusion effect in gas
transport equations, which is based on a diffusion correction for the
dynamics of a simplified process of gas molecule interaction to match
the realistic gas dynamics in the homogenization time scale
limit~\cite{KelMel,Kif,PavStu,Van}. We computed the estimates of the
corrective mass diffusion coefficient in the Euler, Navier-Stokes and
Grad equations for neon, argon, krypton and xenon under the assumption
that the corresponding simplified interaction model of the gas
molecules is that of randomly colliding hard spheres, based on the
experimental data from~\cite{Hut}. We also computed the estimates of
the corrective mass diffusion coefficient for neon, argon, nitrogen,
oxygen, carbon dioxide and methane for the Lennard-Jones random
deflection model of simplified interactions, based on the theoretical
estimates~\cite{HirBirSpo,HirBirSpo2} and experimental
data~\cite{Winn}.  We observed that the available experimental data
seems to support the proposed theory, and that the empirical values of
the corrective diffusion coefficient we used for argon and nitrogen in
the preceding works~\cite{Abr13,AbrOtt} were not too different from
the corresponding theoretical estimates we computed above.

Note that the formula for the corrective diffusion coefficient,
proposed above, is formally valid only when the Schmidt number of the
realistic gas is smaller than that of the chosen simplified
interaction process (otherwise, the diffusion term in the Boltzmann
equation~\eqref{eq:boltzmann} becomes negative). This condition, in
turn, requires that, for the same stress damping rate, the velocity
time series of the real gas exhibit longer autocorrelation times than
those of the simplified random interaction process. The available
experimental data~\cite{Winn} suggests that this condition indeed
holds for various gases, both monatomic and polyatomic, across a wide
range of temperature values. However, we are not currently aware of a
related theoretical result which indicates that the autocorrelation
functions of a random jump process necessarily decay faster than those
of a deterministic deflection process upon which the jump kernel is
constructed, given that their stress correlation functions decay at
equal rate. It would be interesting to see if such result can be
theoretically obtained, either in general, or at least for some
particular examples of multimolecular deflection processes.

It is likely that a similar diffusion correction approach can also be
adopted for more complicated particle dynamics, such as, for example,
electrically charged (ionized) particles and plasma, where the
long-range Coulomb interaction potential is present in addition to the
Lennard-Jones potential. Observe that the proposed theory relies only
on the measured diffusion coefficient and viscosity of a real particle
system, and a suitable random kernel of the deflection approximation
for a given potential, for which the integrals of relevant
autocorrelation functions can be evaluated (even if
approximately). Thus, a similar diffusion correction formalism could
likely apply to the corresponding transport equations.

\ack The author thanks Prof. Jason M. Reese for providing useful
information on the extended fluid dynamics. The work was supported by
the Office of Naval Research grant N00014-15-1-2036.

\end{document}